\documentclass[epj]{svjour}
\usepackage{epsf}
\usepackage[dvips]{graphics}

\begin{document}

\title{ Information and Topology in
Attractor Neural Network }
\titlerunning{ Information and Topology in
Attractor Neural Network }

\authorrunning{ D.Dominguez,
K.Koroutchev,
E.Serrano,
F.B.Rodr\'{\i}guez 
}

\author{ David Dominguez
\thanks
{\emph{ DD. E-mail: david.dominguez@uam.es 
}}	     
Kostadin Koroutchev,
Eduardo Serrano and 
Francisco B. Rodr\'{\i}guez 
}
\offprints{}
\institute{EPS, Universidad Autonoma de Madrid, 
Cantoblanco, Madrid, 28049, Spain 
}

\date{Received:\today }


\abstract{
A wide range of networks,
including small-world topology, 
can be modelled by the connectivity $\gamma$,
and randomness $\omega$ of the links.
Both learning and attractor abilities of a neural network
can be measured by the mutual information (MI),
as a function of the load rate and overlap
between patterns and retrieval states.
We use MI to search for the optimal topology,
for storage and attractor properties of the network.
We find that,
while the largest storage implies an optimal 
$MI(\gamma,\omega)$ at $\gamma_{opt}(\omega)\to 0$,
the largest basin of attraction leads to an
optimal topology at moderate levels of $\gamma_{opt}$,
whenever $0\leq\omega<0.3$.
This $\gamma_{opt}$ is related to the clustering and 
path-length of the network.
We also build a diagram for the dynamical phases with
random and local initial overlap,
and show that very diluted networks lose their attractor ability.
}
\maketitle

\section{Introduction}

The interest on attractor neural networks (ANN),
originally dealing with fully-connected architectures,
has been renewed with the study of more realistic
topologies \cite{MA04},\cite{To04}.
Among them,
the small-world (SW) graph \cite{AB02}, \cite{MA04},
modelled by only two parameters:
$\gamma\equiv K/N$,
the average rate of links, $K$,
per network size, $N$;
and $\omega$, which controls the rate of random links
(among all $K$ neighbors),
can capture most facts of a wide range of networks \cite{SW98}.
The load rate $\alpha=P/K$
(where $P$ is the number of independent patterns).
and the overlap $m$ between neuron states and memorized patterns
are the most used measures of the retrieval ability
of the networks \cite{AG87}.

The overlap as a function of $\alpha$ is plotted in 
upper panels of Fig.\ref{im,ac},
for fully-connected (FC, left panel),
moderately-diluted (MD, central) and extremely-diluted (ED, right)
networks.
The FC network has a critical
$\alpha^{FC}_c\sim 0.138$ \cite{AG87},
with the overlap $m^{FC}_c\sim 0.97$,
and a sharp transition to $m\to 0$ for larger $\alpha\geq\alpha_c$,
where it fails to retrieve,
as seen in the left panel.
However, for ED networks ($K\ll N$),
the transition is smooth.
In particular, the random ED network
(RED $\omega=1.0$, circles and dashed line)
has $\alpha^{RED}_c\sim 0.64$\cite{CG88}
but the overlap falls continuously to $m^{RED}_c\sim 0$,

Less attention has been paid to the
study of the mutual information (MI) between stored patterns
and the neural states \cite{Ok96},\cite{DB98}.
The lower panels of Fig.\ref{im,ac} display the information rate, 
$i$, evaluated from the conditional probability of neuron states
$\vec{\sigma}$ given the patterns $\vec{\xi}$,
for the mean-field (MF) networks we deal with.
The FC case shows a critical information of about
$i^{FC}_c\sim 0.132$.
The RED networks has null information at $\alpha_c$,
$i^{RED}_c\sim 0.0$.
However,
if one look for the value of
$\alpha_{max}$ corresponding to the maximal information
$i_{max}\equiv i(\alpha_{max})$,
instead of $\alpha_c$,
one finds $i_{max}^{RED}\sim 0.223$ for
$\alpha_{max}^{RED}\sim 0.32$.
The FC network has the same $i_{max}^{FC}\equiv i_{c}^{FC}$.

We address the problem of searching the topology which
maximizes the MI.
Using the graph framework,
we built networks with the parameters:
connectivity rate, $\gamma$,
running from the FC ($\gamma=1$) to the ED ($\gamma\to 0$
networks;
and randomness, $\omega$,
ranging from local ($\omega=0$) to random ($\omega=1$) neighbors.
Diluted topologies with $\omega\sim 0.1$,
with large clustering coefficient (C) and 
small mean-path-length (L) between neurons,
so-called small-world (SW),
are rather usefull when one needs fast and robust to noise
information transmition,
without spending too much wiring \cite{MA04}, 
SW networks may model many biological systems,
for instance,
in a brain local connections dominate in intracortex,
while there are a few intercortical connections \cite{RT04}.


\begin{figure*}[t]
\begin{center}
\epsfxsize 15.cm \epsfysize 10.cm
\epsfbox{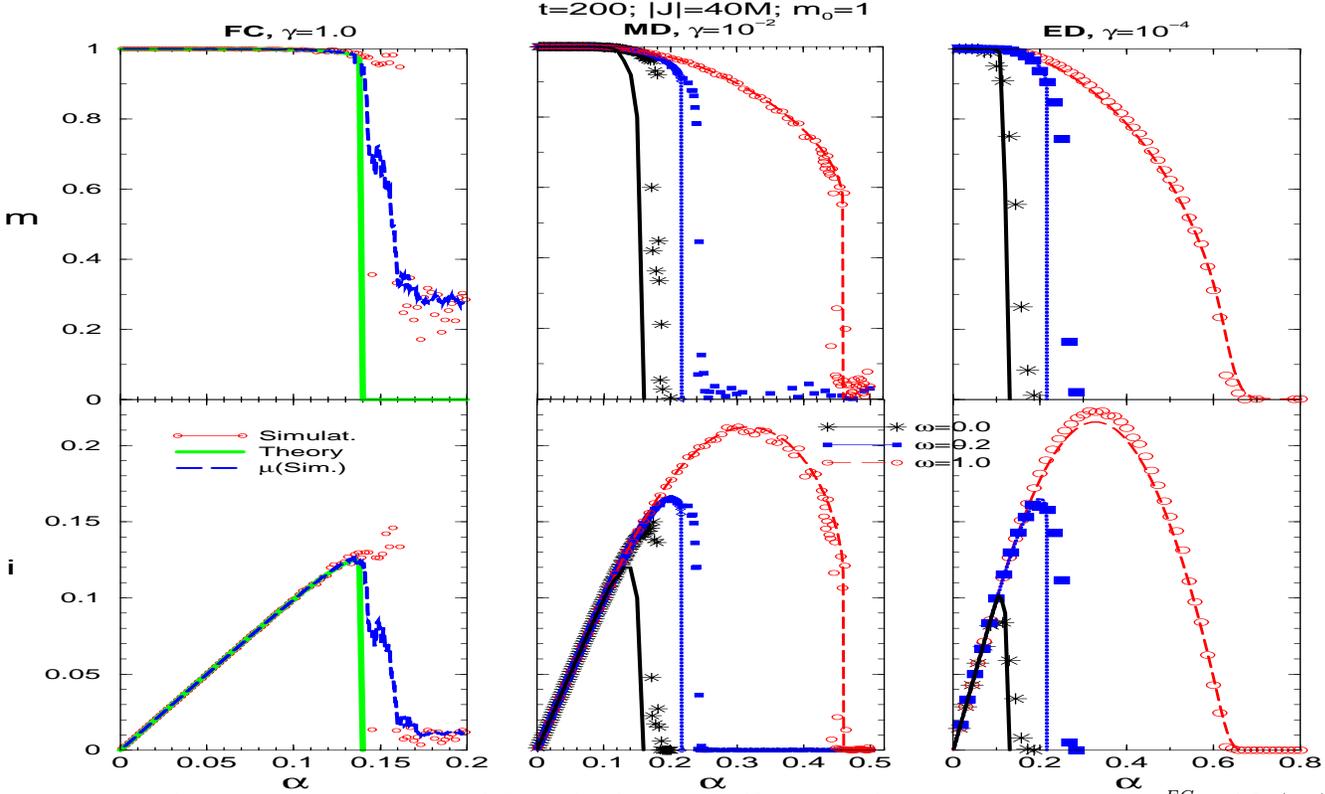}
\caption{ {\it \small{
The overlap $m$ and the information $i$ vs $\alpha$ for
different architectures:
fully-connected, $\gamma^{FC}=1.0$ (left),
moderately-diluted, $\gamma^{MD}=10^{-2}$ (center)
and extremely-diluted, $\gamma^{ED}=10^{-4}$ (right).
Symbols represents simulation with initial overlap $m^0=1$
and $|J|=40M$, with
local (stars, $\omega=0.0$),
small-world (filled squares, $\omega=0.2$),
and random (circles, $\omega=1.0$) connections.
Lines are for theoretical results:
solid, $\omega=0.0$,
dotted, $\omega=0.2$,
and dashed, $\omega=1.0$.
In left, dashed line means averaging the simulation.}
} }
\label{im,ac}
\end{center}
\end{figure*}


The right panels of Fig.\ref{im,ac} plot also $m$ and $i(\alpha)$
for a SW ED network
(SED, $\omega=0.2$), with $i_{max}^{SED}=0.165$,
and for the local ED network
(LED, $\omega=0.0$), with $i_{max}^{LEC}=0.0855$,
it shows how the information increases with randomness $\omega$.
The central panel of Fig.\ref{im,ac}
plot MD networks.
Comparing different dilution levels,
one see that $i$ increases (decreases) with $\gamma$
for local (random) networks,
and remains about the same for SW topologies.
A question arises about the optimal topology:
if the randomness $\omega$ is fixed (by physical constraints),
which is the best connectivity $\gamma$?
To our knowledge, up to now,
no previous answer to this question were known.
We approach this problem from two scenarios:
the stability and the retrieval attractor.
We will show that, 
concerning the stability of a pattern,
the RED network performs the best, $\gamma_{opt}\to 0$.
However, regarding the attractor basins,
the optimal topology holds for MD, 
for instance, $\omega\sim 0.1$ leads to an optimal
$\gamma_{opt}\sim 10^{-2}$.

The structure of the paper is the following:
in the section 2,
we define the topology and neural-dynamics model,
and review the information measures
used in the calculations. 
The results are shown in Sec.3, 
where we study retrieval by theory and
simulation.
We present a diagram for the phases with local and random 
initial conditions,
and show a relation between topology and MI.
Conclusions are drawn in last section.

\section{The Model}

\subsection{Topology and Dynamics}

The synaptic couplings are 
$J_{ij}\equiv C_{ij}W_{ij}$,
where ${\bf C}$ is the topology matrix ${\bf C}$ 
and in {\bf W}  are the learning weights.
The topology splits in local and random links,
$\{C_{ij}= C^l_{ij}+C^r_{ij} \}$.
The local part connects the $K_l$ nearest neighbors,
$C^l_{ij}=\sum_{k\in V} \delta(i-j-k)$,
with $V=\{1,...,K_l\}$ in the asymmetric case,
in a closed ring.
The random part consists of independent random variables 
$\{C^r_{ij}\}$, 
distributed with probability 
$p(C^r_{ij}=1)=c_r$, and $C^r_{ij}=0$ otherwise,
with $c_r=K_r/N$, where $K_r$ is the mean number of
random connections of a single neuron.
Hence, the neuron connectivity is
$K=K_l+K_r$.
The network topology is then characterized by two parameters:
the {\it connectivity} ratio, defined as $\gamma=K/N$,
and the {\it randomness} ratio, $\omega=K_r/K$.
The symmetry constraints seems to play only side effects on the
information properties.
The $\omega$ plays the role of rewiring probability
in the {\em small-world} model (SW) \cite{SW98}.
Our model was proposed by Newman and Watts \cite{NW99},
which has the advantage of avoiding disconnecting the graph.

Note that the topology  ${\bf C}$ can be defined by an 
adjacency list connecting neighbors, 
$i_k, k=1,...,K$, with $C_{ij}=1:j=i_k$.
So the storage cost of this network is $|{\bf J}|=N\cdot K$.
The learning algorithm updates {\bf W},
according to the Hebb rule
\begin{equation}
W_{ij}^{\mu}= W_{ij}^{\mu-1} + 
\xi_{i}^{\mu} \xi_{j}^{\mu}.
\label{2.Ki}
\end{equation}
The network starts at $W^{0}_{ij}=0$,
and after $\mu=P=\alpha K$ learning steps,
it reaches a value 
$W_{ij} = 
\sum_{\mu}^{p} \xi_{i}^{\mu} \xi_{j}^{\mu}$.
The learning stage is a slow dynamics,
being stationary in the time scale of the 
much faster retrieval stage,
we define in the following.


The neural states, $\sigma_i^t\in\{\pm 1\}$,
are updated according to the 
dynamics:
\begin{eqnarray}
\sigma^{t+1}_i={\rm sign}(h_i^t), \,\,
h_{i}^t \equiv \sum_{j} J_{ij} \sigma_j^t ,\,\,i=1...N
\label{2.st}
\end{eqnarray}
In the case of symmetric synaptic couplings, $J_{ij}=J_{ji}$,
an energy function 
$H_{s} = -\sum_{(i,j)} J_{ij} \sigma_i \sigma_j$
can be defined,
whose minima are the stable states of the dynamics 
Eq.(\ref{2.st}).

In the present paper, 
we work out the asymmetric network by simulation
(no constraints $J_{ij}=J_{ji}$).
The theory was carried out for symmetric networks.
Biological networks are usually asymmetric \cite{RT04},
but this feature does not allow any thermodynamics approach. 
As it is seen in Fig.\ref{im,ac},
theory and simulation shows similar results,
except for local networks 
(theory underestimate $\alpha_{max}$),
where the symmetry may play some role. 
A stochastic macro-dynamics takes place due to the extensive 
learning of $P=\alpha K$ patterns.

\subsection{The Information Measures}


The network state at a given time $t$ is defined by 
a set of binary neurons,
$\vec{\sigma}^{t}=\{\sigma_i^t\in\{\pm 1\},i=1,...,N\}$.
Accordingly, each pattern
$\vec{\xi}^{\mu}=\{\xi_i^{\mu}\in\{\pm 1\},i=1,...,N\}$,
is a set of site-independent random variables,
binary and uniformly distributed:
$p(\xi^{\mu}_{i}=\pm 1)= 1/2$.
The network learns a set of independent patterns
$\{\vec{\xi}^{\mu},\,\,\mu=1,...,P\}$.

The task of the neural channel is to retrieve a pattern 
(say, $\vec{\xi}^{} \equiv \vec{\xi^{\mu}}$) 
starting from a neuron state 
$\vec{\sigma}^{0}$ which is 
inside its attractor basin.
This is achieved through a network dynamics
coupling neurons 
through a {\em synaptic matrix} 
${\bf J}\equiv\{J_{i,j}\}$
with cardinality $|{\bf J}|=N\times K$.
The relevant order parameter is the $overlap$ between the
neural states and the pattern:
\begin{equation}
m^{t}_{N}\equiv
{1\over N}\sum_{i}\xi^{}_{i}\sigma_{i}^t,
\label{3.mm}
\end{equation}
at the time step $t$.
Together with the overlap,
one needs a measure of the load,
which is the rate of pattern bits per synapses used to store them.
Since the synapses and patterns are independent,
the load is given by 
$\alpha=|\{\vec{\xi}^{\mu}\}|/|{\bf J}|\equiv P/K$.

We require the interactions ${\bf J}$ to be long-range,
and neglect spatial correlation.
Hence, we regard a mean-field network (MFN),
the distribution of the states is assumed to be 
site-independent.
Therefore,
according to the law of large numbers,
the overlap can be written, 
for $K,N\to\infty$, as 
$m^t= \langle
\sigma^t{\xi}
\rangle_{\sigma,\xi}$.   
The brackets represent average over the joint distribution 
$p(\sigma,\xi)$, 
for a single neuron,
understood as an ensemble distribution
for the neuron states $\{\sigma_{i}\}$ 
and pattern $\{\xi_{i}\}$ \cite{DB98}.


This distribution factorizes in the conditional probability
$p(\sigma|\xi)=(1+m\sigma\xi)\delta(\sigma^2-1)$, \cite{BD00}
and input probability $p(\xi)$.
In $p(\sigma|\xi)$,
all types of noise in the retrieval process are enclosed
(both from environment and over the dynamical process itself) .
With the above expressions and
$p(\sigma)\equiv\delta(\sigma^2-1)$,
we can calculate the MI \cite{DB98},
a quantity used to measure the prediction that an observer 
at the output ($\bf{\sigma}$) can do about the input 
($\bf{\xi}^{\mu}$) (we drop the time index $t$).
It reads 
$MI[\sigma;\xi]=S[\sigma]-S[\sigma|\xi]$,
where the output and conditional entropies are given
(in bits) by \cite{BD00}:
\begin{eqnarray}
&&
S[\sigma|\xi]
= 
-{1+m\over 2}\log_2{1+m\over 2} - 
{1-m\over 2}\log_2{1-m\over 2} , \,\,
\nonumber\\
&& S[\sigma]=1[bit].
\label{3.He}
\end{eqnarray}

We define the information rate as 
\begin{equation}
i(\alpha,m)=MI[\vec{\sigma}|\{\vec{\xi}\mu\}]/|{\bf J}|\equiv
\alpha MI[\sigma;\xi],
\label{3.ia}
\end{equation}
since for independent neurons and patterns,
$ MI[\vec{\sigma}|\{\vec{\xi}\mu\}] \equiv
\sum_{i\mu} MI[\sigma_i|\xi^{\mu}_{i}] $.
The information is $i=\alpha MI$, Eq.(\ref{3.ia}),
where the load rate is scaled as $\alpha=P/K$.

When the network approaches its saturation limit $\alpha_c$,
the neuron states can not remain close to the patterns,
then $m_c$ is usually small.
So, while the number of patterns increases,
the information per pattern decreases.
Therefore,
information $i(\alpha,m)$ is a non-monotonic function of
the overlap and load rate 
(see Fig.\ref{im,ac}), 
which reaches its maximum value $i_{max}=i(\alpha_{max})$
at some value $\alpha_{max}\leq\alpha_{c}$
of the load.

\section{Results}


\begin{figure}[t]
\begin{center}
\epsfxsize 8.cm \epsfysize=10.cm
\epsfbox{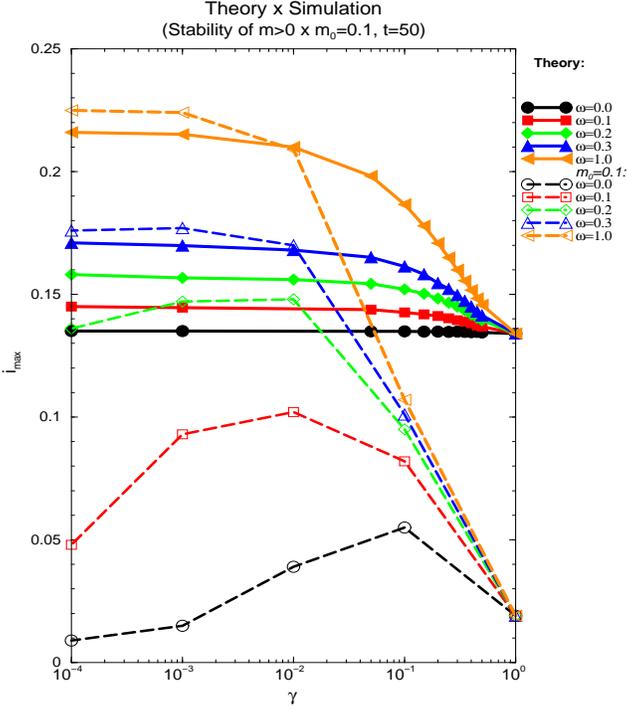}
\caption{\small
Maximal information $i_{max}=i(\alpha_{max})$ vs $\gamma$.
Theory for the stationary states (solid lines),
and simulations with $m^0=0.1$ (dashed lines),
with several values of randomness $\omega$.}
\label{i,gwt}
\hfill
\end{center}
\end{figure}

We studied the information for the stationary and dynamical states 
of the network,
as a function of the topological parameters,
$\omega$ and $\gamma$.
A sample of the results for simulation and theory is shown
in Fig.\ref{im,ac},
where the stationary states of the overlap and information are plotted
for the FC, MD and ED architectures.
It can be seen that the information increases with dilution and with
randomness of the network.
A reason for this behavior is that dilution decreases the correlation
due to the interference between patterns.
However, dilution also increases the mean-path-length of the network,
thus, if the connections are local,
the information flows slowly over the network.
Hence, the neuron states can be eventually trapped in noisy patterns.
So, $i_{max}$ is small for $\omega\sim 0$ even if $\gamma=10^{-4}$.

\subsection{Theory: Storage}

The theoretical approach follows the Gardner 
calculations\cite{CG88}.
A supposition is that the network state is near a given pattern.
At temperature T=0 the MF approximation gives
the fixed point equations:
\begin{eqnarray}
  && m = {\rm erf}(m/\sqrt{r\alpha}),\,\,  \\
  && \chi = 2\varphi(m/\sqrt{r\alpha})/\sqrt{r\alpha};\\
  && r = \sum_{k=0}^\infty a_k (k+1) \chi^{k},
  \,\;a_k=\gamma Tr[({\bf C}/K)^{k+2}]
\label{4.me}
\end{eqnarray}
with ${\rm erf}(x)\equiv 2\int_0^{x} \varphi(z) dz $,
$\varphi(z)\equiv e^{-z^2/2}/\sqrt{2\pi} $.
The parameter $a_k$ 
is the probability of existence
of cycle of length $k+2$ in the graph ${\bf C}$.
The $a_k$ can be calculated either by using Monte Carlo \cite{DK04},
or by an analytical approach,
which gives $a_k\sim \sum_m \int d\theta [p(\theta)]^k e^{im\theta}$,
where $p(\theta)$ is the Fourier transform of the probability of
links, $p(C_{ij})$.
For an RED and FC networks one recover the known results
for $r^{RED}=1$ and $r^{FC}=1/(1-\chi)^2$ respectively \cite{HK91}.

The theoretical dependence of the information on the load,
for FC, MD and ED networks,
with local, small-world and random connections, 
are plotted in the solid lines in Fig.\ref{im,ac}.
A comparison between theory and simulation is also given
in Fig.\ref{im,ac}.
It can be seen that both results agree for most $\omega>0$,
but theory fails for $\omega=0$.
One reason is that theory uses symmetric constraint,
while simulation was carried out with asymmetric synapsis.
The solid lines in Fig.\ref{i,gwt} shows their maxima 
$i_{max}(\gamma,\omega)\equiv i(\alpha_{max},\gamma,\omega)$ vs. 
the parameter $\omega$, varying $\gamma$. 
It is seen that thermodynamical optimum $i(\gamma_{opt})$ is at 
$\omega_{opt}\to 1, \gamma\to 0$.
This implies that the best topology for information,
respect to the stationary states,
is the RED network.
It is worth to note that the simulation converges to
the theoretical results if $m_0=1.0$ when $t\to\infty$,
this means that theory accounts for the storage capacity of
the network.
However, 
quite different qualitative behavior holds for the simulation 
with low $m_0=0.1$,
see Fig.\ref{i,gwt}.
displays optima $i(\gamma_{opt})$ for MD topologies.

\subsection{Simulation: Attractors}

The theoretical equations for the stationary states,
Eqs.(\ref{4.me}),
account only for the existence of the retrieval (R)
solution $m>0$.
However, they say nothing about its stability.
The zero states (Z), $m=0$, 
are also a solution of Eqs.(\ref{4.me}),
so both R and Z may coexist in some region of topological
parameters $\gamma$, $\omega$.
In order to study the stability of the attractors,
we simulated Eq.(\ref{2.st}),
and check how the network behaves for different initial conditions.
 
Both local and random connections are asymmetric.
The simulation was carried out with 
$N\times K=36\cdot 10^{6}$ synapses,
storing an adjacency list as data structure,
instead of $J_{ij}$.
For instance, with $\gamma\equiv K/N=0.01$,
we used $K=600,N=6\cdot 10^{4}$.
In \cite{MM03} the authors use $K=50,N=5\cdot 10^{3}$,
which is far from asymptotic limit.
We averaged over a window in the axis of $P$.

\begin{figure*}[t]
\begin{center}
\epsfxsize 15.cm 
\epsfysize 10.cm
\epsfbox{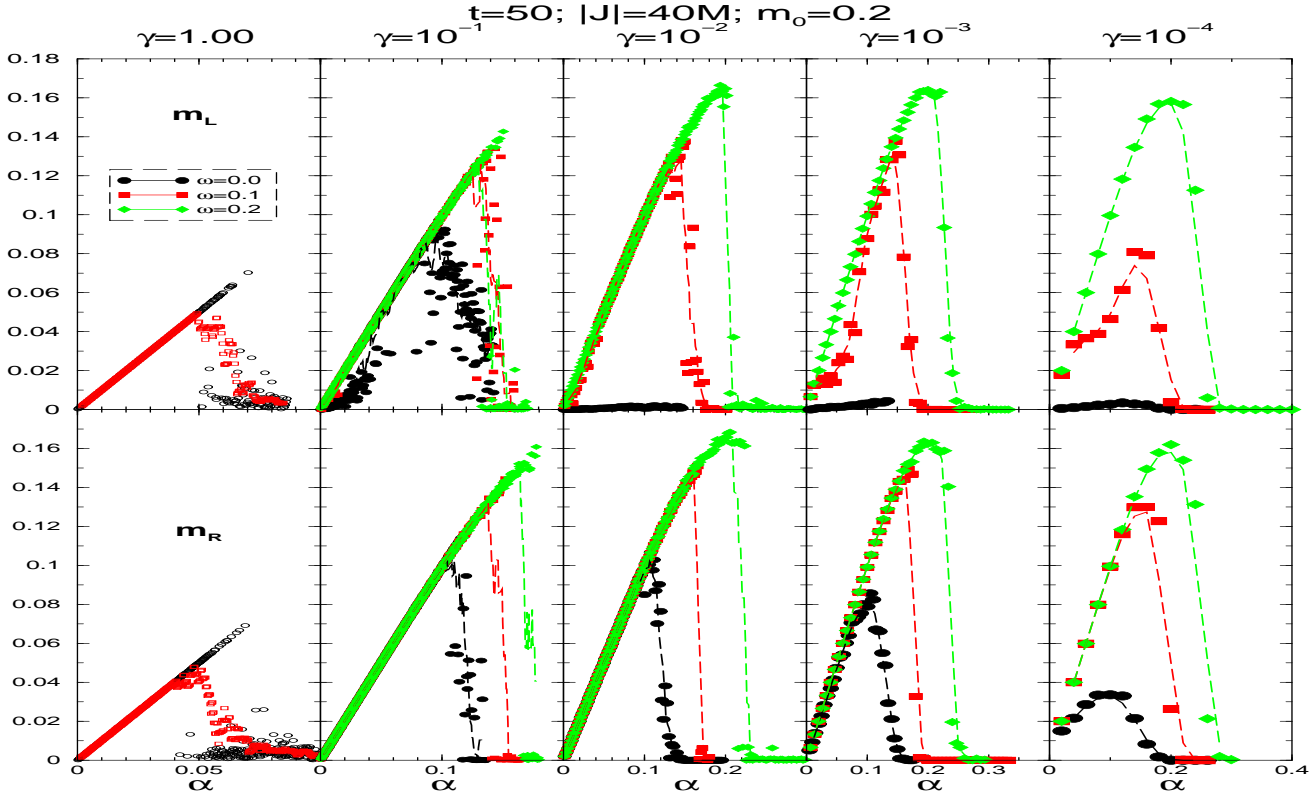}
\caption{\small 
Information $i(\alpha)$ for initial conditions:
$m_r$ and $m_l$ (always $m^0=0.2$).
Simulations with $N.K=4.10^7$,
for $\gamma=1,10^{-1},10^{-2},10^{-3},10^{-4}$ (from left to right) and 
$\omega=0.0, 0.1, 0.2$ (circles, squares, diamonds).
and several $\omega$. }
\label{i,agw}
\end{center}
\end{figure*}

To check for the attractor properties of the retrieval,
the neuron states start far from a learned pattern,
but inside its basin of attraction,
$\vec{\sigma}^{0}\in B(\vec{\xi}^{\mu})$.
First, we choose an initial configuration
given by a random correlation with patterns,
$p(\sigma^0=\pm\xi^{\mu}|\xi^{\mu})=(1\pm m^0)/2$,
for all neurons
(so we avoid a bias between local/random neighbors).
We call this the $m_R$ initial overlap.
The retrieval dynamics starts with an overlap $m^{0}=0.2$,
and stops after $t_f= 50$ steps 
(unless it converges to a fixed point $m^*$ before $t=t_f$).
Usually, 
$t_f=20$ parallel (all neurons) updates
is a large enough delay for retrieval.
The information $i(\alpha,m;\gamma,\omega)$
is calculated.
We averaged over a window in the axis of $P$,
usually $\delta P=25$.

The results are depicted in the dashed curves of 
Fig.\ref{i,gwt},
where the $i_{max}(\gamma,\omega)$ are plotted against
$\gamma$.
One see that, 
unlike the theoretical results for the storage capacity,
there are MD topologies which performs the 
best when the attractor properties are considered.
Starting with $m_0=0.1$ yields optima $i(\gamma_{opt})$
for moderate dilutions,
for instance, 
with $\omega=0.2$, 
it holds $\gamma_{opt}\sim 10^{-2}$.

Next, we compare $m_R$ with another type of initial 
distribution.
The neurons start with local correlations:
$\sigma^0_i=\xi^{\mu}_i,\,i=1,...,(Nm^0)$,
and random $\sigma^0_i$ otherwise.
We call it the $m_L$ initial overlap.
The results are shown in the Fig.\ref{i,agw}.
In the lower (upper) panels we see the behavior with the 
$m_R$ ($m_L$) initial overlap. 
The first observation now is that the maxima information 
$i_{max}(\gamma;\omega)$ increases with dilution 
(smaller $\gamma$) if the network is more random,
$\omega\simeq 1$,
while it decreases with dilution if the network is more local,
$\omega\simeq 0$.
However, 
there is a moderate $\gamma_{opt}$ for which the 
information $i(\gamma_{opt})$ is optimized.
For instance with $\omega=0.1$,
starting with initial overlap $m_R$, 
the optimum is $i(\gamma_{opt})\sim 0.148$ at 
$\gamma_{opt}=10^{-2}$.
For $m_L$,
the optimum is $i(\gamma_{opt})\sim 0.138$ at 
$\gamma_{opt}=10^{-2}$.
We see that the initial $m_R$ allows for an easier retrieval
for any $\omega$,
but local topologies ($\omega=0$)
are very sensitive to the type of initial overlap,
and lose their retrieval abilities if the connectivity is
$\gamma_{opt}\leq 10^{-2}$.
We also observe in Fig.\ref{i,agw} a feature of the 
$m_L$ condition:
the network improves its retrieval ability with learning
($m$ increases with $\alpha$) 
before the information reaches its maxima,
which resembles a stochastic resonance effect.

\begin{figure}[t]
\begin{center}
\epsfxsize 8.cm 
\epsfysize 8.cm
\epsfbox{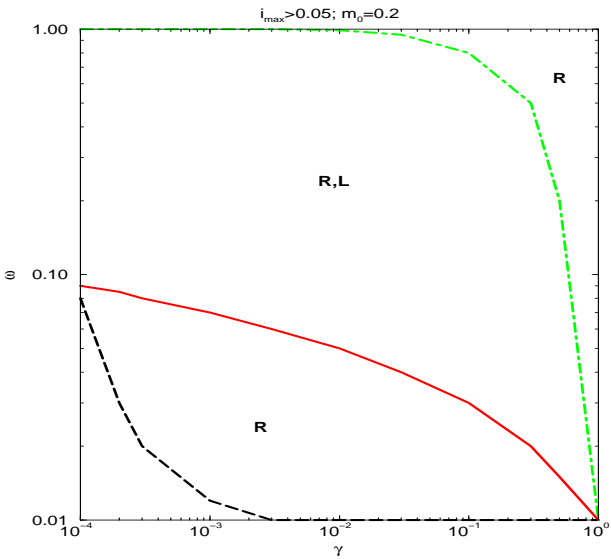}
\caption{\small 
Diagram ($\omega\times\gamma$) with the phases R and L,
for initial overlap $m_0=0.2$. }
\label{w,cm2}
\end{center}
\end{figure}

The comparison between upper ($m_L$) and lower ($m_R$)
panels of Fig.\ref{i,agw},
shows that the non-monotonic behavior of the information with
dilution,
is stronger for the local than for the random initial overlap.
This sensitivity to the initial conditions 
can be understood in terms of the basins of attraction.
Random topologies have very deep attractors,
specially if the network is diluted enough,
while regular topologies almost lose their retrieval abilities
with dilution.
However, since the basins becomes rougher with dilution,
then network takes longer to reach the attractor,
and can be trapped in metastable states.
Hence, the competition between depth-roughness is won by
the more robust MD networks.

The retrieval capability of the network when start at condition
$m_R$ or $m_L$ is plotted in Fig.\ref{w,cm2}.
We represent as $R$ the phase where the retrieval reaches 
at least the information $i_{max}=0.05$,
starting from $m_0=0.2$, with $m_R$.
The phase $L$ is the same, 
but starting with $m_L$.
Good retrieval ($i_{max}\geq 0.05$) is not allowed with the 
$m_L$ condition neither for very connected nor for 
local diluted topologies.

\subsection{Clustering and Mean-Length-Path}

\begin{figure*}[t]
\begin{center}
\epsfxsize 15.cm
\epsfysize 10.cm
\epsfbox{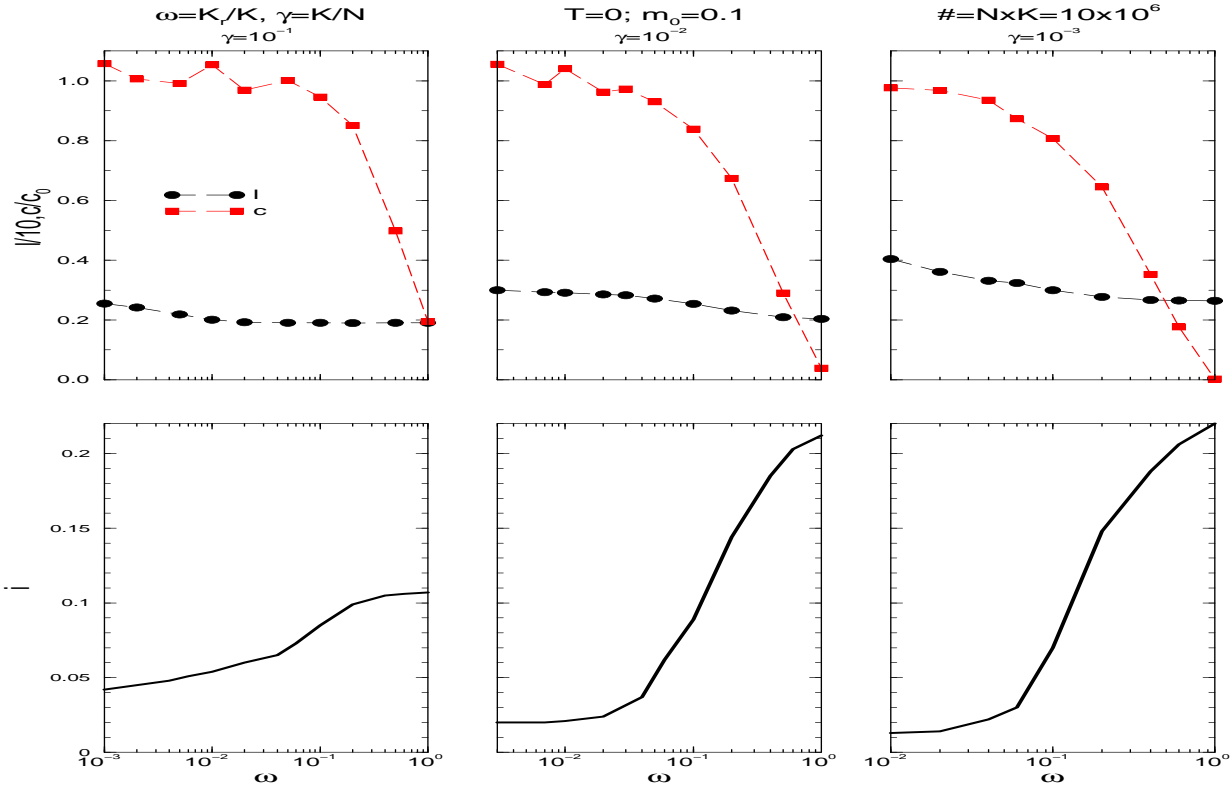}
\caption{\small
The maxima information $i(\alpha_{max})$ (bottom),
clustering coefficient $c$ 
and mean-path-length $l$ (top) vs $\omega$,
for $\gamma=10^{-1},10^{-2},10^{-3}$ (from left to right).
Simulations with  $N.K=10M$. }
\label{icl,w}
\end{center}
\end{figure*}

We described here the topological features of the network,
as a function of its parameters,
the 
clustering coefficient, $c$,
and the mean-length-path between neurons, $l$.
When $\gamma$ is large,
the net has $c$ large, $c\sim 1$ and  
$l$ small, $l=O(1)$, 
whatever $\omega$ used.
When $\gamma$ is small,
then if $\omega\sim 0$, 
the net is clustered ($c=O(\gamma)$), 
and has large paths ($l\sim N/K$);
if $\omega\sim 1$,
the net becomes random
($c\ll 1$ and $l\sim \ln N$).
However, 
if the randomness is about $\omega\sim 0.1$,
then $c=O(\gamma)$ but with $l\sim \ln N$,
and the network behaves as a small-world (SW):
clustered but with short paths.

The dependence of $c$, $l$ and $i$ with randomness $\omega$ 
are plotted in Fig.\ref{icl,w}.
In all panels, for connectivity 
$\gamma = 10^{-1}$, $10^{-2}, 10^{-3}$,
we see a decrease of $c$ and $l$ and an increase of $i$
with $\omega$.
For these range of $\omega$,
the path-length $l$ has already decreased to small values,
and the networks have entered the SW region.
However, in the right panels,
there is still some slow down of $l$.
The clustering $c$ decreases fast around $\omega=0.2$,
after which the network is random-like.

This region $0.001\leq\omega\leq 0.2$ is the SW regime for the 
network we study.
On the other hand,
looking at the bottom panel,
we see that at the end of the SW graph,
between $0.05\leq\omega\leq 0.20$
there is a fast increase of $i$. 
We conjecture that after the SW region,
a further increase of the randomness $\omega$ does not worth 
its wiring cost to gain a little extra information $i$.

\section{Conclusions}

We have discussed the behavior of the information capacity
of an attractor neural network with the topology.
We calculated the mutual information for a Hopfield model,
with Hebbian learning,
varying the connectivity ($\gamma$) and 
randomness ($\omega$) parameters,
and obtained the maximal respect to $\alpha$,
$i_{max}(\gamma,\omega)\equiv i(\alpha_{max};\gamma,\omega)$.
The information $i_{max}$ always increase with $\omega$,
but for a fixed $\omega$,
an optimal topology $\gamma_{opt}$,
in the sense of the information,
$i_{opt}\equiv i_{max}(\gamma_{opt},\omega)$,
can be found.
We presented stationary and attractor states.

From the stability calculations,
the optimal topology respect to the storage,
is the random extremely diluted (RED) network.
Indeed, if no pattern completion is required,
they can be stored statically,
and the best way is without any connectivity.  
For retrieval dynamics, however, this is not true:
we found there is an intermediate $\gamma_{opt}$,
for any fixed $0\leq\omega<0.3$.
Moreover, 
local diluted (LD) networks are even more damaged
if they starts with local overlap 
than if the initial overlap is random.
This can be understood regarding the shape of the attractors.
The ED waits much longer for the retrieval than
more connected networks do,
so the neurons can be trapped in spurious states with
vanishing information.
We found there is an intermediate optimal $\gamma_{opt}$,
whenever $0\leq\omega<0.3$.

We found a relation between the fast increase of information 
with $\omega$ and the region of small-world of the topology.
This implies that it worth the wiring cost to stay at the end 
of the SW zone. 
Both in nature and in technological approaches to neural devices,
dynamics is an essential issue for information process.
So, an optimized topology holds in any practical purpose,
even if no attemption is payed to wiring or other energetic costs
of random links \cite{MK04}.
The reason for the intermediate $\gamma_{opt}$ is a competition 
between the broadness
(larger storage capacity)
and roughness (slower retrieval speed)
of the attraction basins.

We believe that the maximization of information respect
to the topology could be a biological criterion
(where non-equilibrium phenomena are relevant)
to build real neural networks.
We expect that the same dependence should happens for
more structured networks and learning rules.
More complex initial conditions may also play some rule in the
retrieval and it worth a dedicated study.

{\bf Acknowledgments}
Work supported by grants TIC01-572, TIN2004-07676-C01-01,
BFI2003-07276,
TIN2004-04363-C03-03
from MCyT, Spain.
DD thanks a Ramon y Cajal grant from MCyT.


\end{document}